\documentclass{aa}
\usepackage{graphics}
\usepackage{txfonts}
\usepackage{afterpage}

\def\bc{\begin{center}}
\def\ec{\end{center}}
\def\be{\begin{equation}}
\def\ee{\end{equation}}
\def\bea{\begin{eqnarray}}
\def\eea{\end{eqnarray}}

\def\mkm{\mu{\rm m}}

\begin{document}
\title{Is the silicate emission feature only influenced by grain size?}

\author{ N.V.~Voshchinnikov\inst{1,2}
         and
         Th.~Henning\inst{3}
         }
\authorrunning{Voshchinnikov and Henning}

\institute{
Sobolev Astronomical Institute,
St.~Petersburg University, Universitetskii prosp. 28,
           St.~Petersburg, 198504 Russia,
e-mail: {\tt nvv@astro.spbu.ru}
\and
 Isaac Newton Institute of Chile, St.~Petersburg Branch
\and
Max-Planck-Institut f\"ur Astronomie, K\"onigstuhl 17, D-69117 Heidelberg, Germany
}
\date{Received March 3, 2008 / accepted March 21, 2008 }

  \abstract{
The flattening of the 10~$\mkm$ silicate emission feature
observed in the spectra  of  T~Tauri and Herbig Ae/Be stars
is usually interpreted as  an indicator of grain growth.
We show in this paper that a similar behaviour of the feature shape occurs
when the porosity of composite grains varies.
We modelled the fluffy aggregates with inclusions of different sizes with
multi-layered spheres  consisting of
amorphous carbon, amorphous silicate, and vacuum.
We also found that
the inclusion of crystalline silicates in composite porous particles
can lead to a shift of the known resonances and production of new ones.

      }
\maketitle

\section{Introduction}

\enlargethispage{\baselineskip}
The shape and strength of the silicate emission feature observed
near 10~$\mkm$ in the spectra of
T~Tauri and Herbig Ae/Be (HAeBe) stars is commonly used as a measure of grain
growth in protoplanetary discs (see Natta et al.~\cite{nattaetal07},
for a review).
It is well-known theoretically that
with an increase of the grain size,
the  feature becomes wider and eventually fades away.
In the case of compact spherical grains with a composition
of astronomical silicate, the 10~$\mkm$  feature disappears when
the grain radius exceeds $\sim 2 \, \mkm$ (see Fig.~\ref{f-asil}).

The standard approach to modelling the 10~$\mkm$  feature
includes calculations of light absorption by several populations of
compact {\rm (and hollow)} silicate spheres.
We used amorphous and crystalline particles  of small and large sizes
to fit the observed emission profiles.
The model was first suggested by Bouwman  et al. (\cite{bou01})
and then modified by van Boekel et al.~(\cite{vb05}).
It was used by Schegerer et al.~(\cite{schetal06});
Honda et al.~(\cite{honda06});
Kessler-Silacci et al.~(\cite{k-s06}); Sargent et al.~(\cite{sar06});
Sicilia-Aguilar et al.~(\cite{aurore07}); and
{\rm Bouwman  et al. (\cite{bou08})}
to fit the observational data.
Further the authors searched for correlations between
the estimated mass fractions of large and crystalline
grains and different stellar
and disc parameters like mass, luminosity, age, spectral type, etc.
As a rule, the correlations are absent or rather weak
(see Table 5, Sicilia-Aguilar et al.,~\cite{aurore07}).

In this Letter, we show that
the shape, position, and strength of the 10~$\mkm$ silicate feature
is also influenced by variations of the properties of
small mass composite aggregate grains.
We modelled the fluffy aggregates
by multi-layered spheres
(see also Voshchinnikov et al.,~\cite{vihd06}).
This particle model permits us to include an arbitrary fraction of
materials, and computations do not require large resources.

\section{Model of porous composite grains}\label{model}

\enlargethispage{\baselineskip}
Fluffy particles should appear as a result of
grain coagulation in interstellar clouds and protoplanetary
discs (e.g., Henning \& Stognienko,~\cite{hs96};
Dominik \& Tielens,~\cite{dt97}; Jones,~\cite{jones04};
Ormel et al.,~\cite{ormel07}).
It is expected that aggregates can consist of several
silicate and carbon sub-particles of different sizes
that can be treated as inclusions in a vacuum matrix.

We use the model of spherical multi-layered particles {\rm introduced} by
Voshchinnikov \&  Mathis~(\cite{vm99}).
Later,  Voshchinnikov et al.~(\cite{vih05}) demonstrated that
the optical properties of layered spheres resemble those of
fluffy aggregates with inclusions of different sizes.

Our model parameters are: the refractive indices and volume fractions
$V_i /V_{\rm total}$ of the materials and the radius of compact particles
$r_{\rm compact}$.
The amount of vacuum in the particle (the {\it particle porosity}
${\cal P}$, $0 \leq {\cal P} < 1$) is
\be
{\cal P} = V_{\rm vac} /V_{\rm total}
= 1 - V_{\rm solid} /V_{\rm total},  \label{por}
\ee
where $V_{\rm solid}$ is the sum of the {\rm volumes} of all species
excluding vacuum. The  radius of porous particles is related to
that of compact particles as
\be
r_{\rm porous} = \frac{r_{\rm compact}}{(1-{\cal P})^{1/3}}
= \frac{r_{\rm compact}}{(V_{\rm solid} /V_{\rm total})^{1/3}}. \label{xpor}
\ee
The model of layered spheres combines all {\rm components},
including vacuum in {\it one}
particle (internal mixing) in contrast to the standard approach
discussed above where ``external mixing''
{\rm (mixture of several individual populations of compact grains)}
is used.

For calculations, we use different kinds of carbon and silicates:
amorphous carbon Be1 and AC1 (Rouleau \& Martin, \cite{rm91}),
amorphous silicate with olivine composition
(Dorschner et al.,~\cite{doretal95}),
crystalline olivine (Fabian et al.,~\cite{fabetal01}),
and astronomical silicate (astrosil; Laor \&  Draine,~\cite{laordr93}).

\section{Analysis of silicate emission}\label{irr}

The silicate feature in the $N$ band was observed in spectra of
a large variety of objects (see Henning~\cite{he07}, for a recent review).
{\rm We should caution the reader that we calculate absorption
efficiencies, but measure the fluxes. A detailed analysis of disc
spectra certainly requires radiative transfer calculations.
Assuming that the silicate emission is optically thin, we
can compare observed fluxes with theoretical profiles. The latter
are proportional to the product of particle absorption cross section
by the Planck function with the particle temperature
$\propto C_{\rm abs}(\lambda)\,B_{\lambda}(T_{\rm d})$.
Radiative transfer calculations show that  grains of
different temperatures contribute to the silicate feature
(see Fig.~1 in Schegerer et al.,~\cite{schetal06}).
However, the dominant contribution for the 10~$\mkm$  feature
comes from  particles with
$T_{\rm d} = 200 - 400$K for which the Planck function is
approximately constant in $N$ band. Therefore, we can adopt
that the shape of the feature depends primarily on the emission
properties of grains\footnote{\rm Note that our case differs from
the case analysed by Li et al.~(\cite{lili04}) where the
cometary particles of different composition are located
at the same distance from the Sun.}.}

The profile of the feature
can be described by the {\it normalised absorption efficiency factor}
$
{Q_{\rm abs}(\lambda)}/{Q_{\rm abs}(\lambda_0)}-1,
$
where the flux at wavelength $\lambda_0$ characterises the continuum.
As usual, the value of $\lambda_0=8.2\,\mu{\rm m}$ is chosen
(e.g., Schegerer et al.,~\cite{schetal06}).

Another representation of the optical behaviour is provided by the
{\it mass absorption coefficient},
which is the ratio of absorption cross section
$C_{\rm abs}(\lambda)$ to particle mass. In the case of a sphere
of porosity ${\cal P}$, it can be written as
\be
\kappa_{\rm abs}(\lambda) = \frac{C_{\rm abs}(\lambda)}
{\rho_{\rm d} V_{\rm total}} =
  \frac{3 \, Q_{\rm abs}(\lambda)}{4\,\rho_{\rm d,\,solid} \,
  r_{\rm s,\, compact} \,(1-{\cal P})^{2/3}} \,,
        \label{kap}
\ee
where $\rho_{\rm d} = \rho_{\rm d,\,solid} (1-{\cal P})$ is the
mean particle density.
The value of $\rho_{\rm d,\,solid}$ is obtained by averaging
the density of all species excluding vacuum.
In our calculations, we assume that $\rho_{\rm d,\,Si}=3.3$\,g/cm$^3$ and
$\rho_{\rm d,\,C}=1.85$\,g/cm$^3$ for silicate and carbon, respectively.

\begin{figure}[ht!]
\bc
\resizebox{\hsize}{!}{\includegraphics{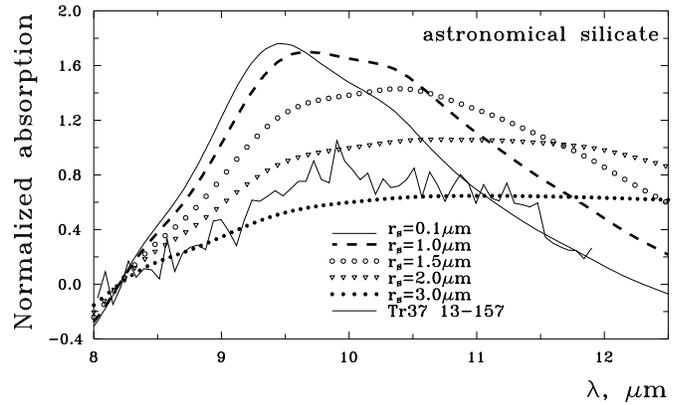}}
\caption{Wavelength dependence of the normalised
absorption efficiency factor
[$Q_{\rm abs}(\lambda)/Q_{\rm abs}(8.2\,\mu{\rm m})-1$]
for compact spheres consisting of astrosil.
The effect of variation of the silicate emission shape with
the particle size is illustrated.
The observational profile of the T~Tauri star
Tr~37~13-157 (Sicilia-Aguilar et al.,~\cite{aurore07}) is shown for comparison.
{\rm Note that this profile is not continuum subtracted.}
}\label{f-asil}\ec
\vspace*{-0.5cm}
\end{figure}

Figure~\ref{f-asil} shows the wavelength dependence of
the normalised absorption efficiency factors for compact silicate
spheres of diverse sizes. With a growth of the particle size (and mass),
the 10~$\mkm$ silicate feature  broadens, its height decreases
and the position of maximum shifts to longer wavelengths.


\begin{figure}[hb!]
\bc
\resizebox{8.6cm}{!}{\includegraphics{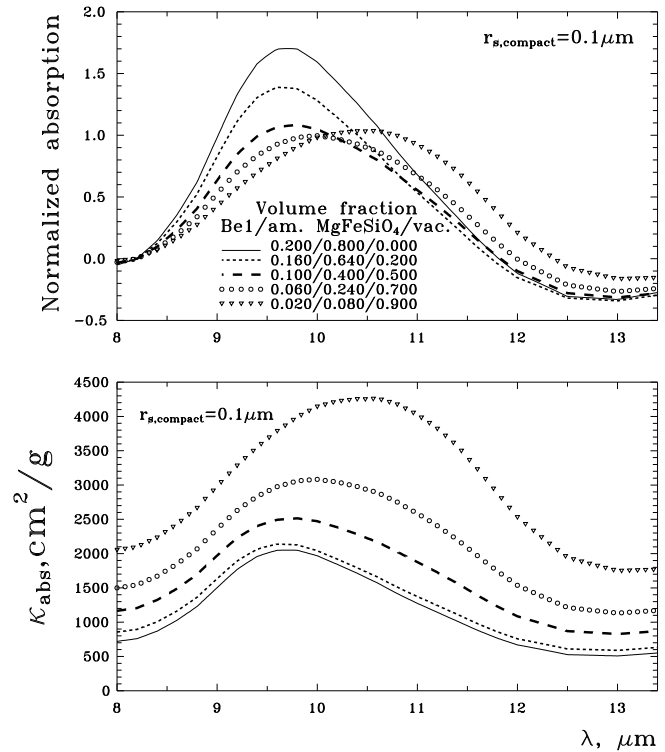}}
\caption{Wavelength dependence of the normalised
absorption efficiency factor (upper panel) and
mass absorption coefficient (lower panel)
for layered spheres with $r_{\rm s,\,compact}=0.1\,\mkm$.
The particles consist of amorphous carbon (Be1),
amorphous silicate with olivine composition (MgFeSiO$_4$), and vacuum.
The volume fraction of components $V_i /V_{\rm total}$ is indicated.
The particles are of the same mass
($V_{\rm Be1}/V_{\rm solid}=V_{\rm Be1}/(V_{\rm Be1}+V_{\rm MgFeSiO_4})=0.2$),
but of different porosity.
The effect of variation of the silicate emission shape with the
particle porosity is illustrated.
}\label{f-ob}\ec
\vspace*{-0.5cm}
\end{figure}

{\it In a similar manner, the silicate feature changes
when the particle porosity grows}
(Fig.~\ref{f-ob}). However, in this case the particle size increases only
moderately, while  particle mass remains the same.
With a growth of porosity, the peak strength decreases
for normalised absorption (Fig.~\ref{f-ob}, upper panel)
whereas the mass absorption coefficient becomes larger
(Fig.~\ref{f-ob}, lower panel).
The value of $\kappa_{\rm abs}$ almost doubles at the peak
position when we replace the compact particle by porous particle.

It is well known that the shape and strength of the silicate feature
depend on the type of the amorphous silicate, particle size, and
fractal dimension
(see, e.g., van Boekel et al.,~\cite{vb05};
Schegerer et al.,~\cite{schetal06}; Min et al.,~\cite{min06}).
Using the model of
composite grains we can also
investigate how carbon embedded in  particles affects  the
characteristics of the silicate feature.


\clearpage

\begin{figure}[ht]
\bc
\resizebox{\hsize}{!}{\includegraphics{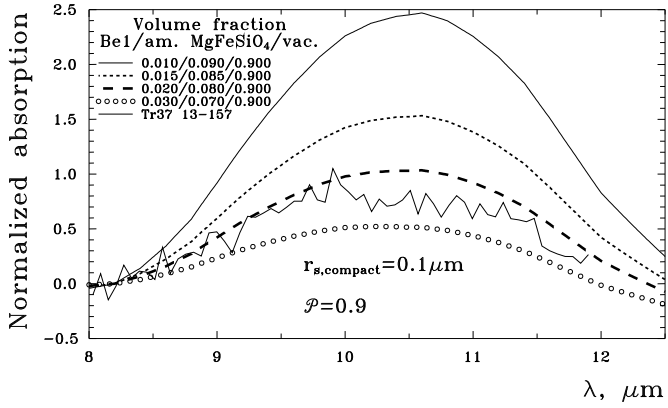}}
\caption{The same as in Fig.~\ref{f-ob} (upper panel) but now
for particles of porosity ${\cal P}=0.9$.
The particles are of the same porosity but of slightly different mass.
The value of
$V_{\rm Be1}/V_{\rm solid}$ varies from 0.1 to 0.3.
The effect of variation of the silicate emission shape with
the volume fraction of carbon is illustrated.
The observational profile of the T~Tauri star
Tr~37~13-157 
is shown for comparison.
}\label{f-09}\ec
\vspace*{-0.5cm}
\end{figure}

\noindent This dependence is plotted
in Fig.~\ref{f-09}.  With addition of carbon, the feature very rapidly
transforms into a plateau and then disappears.


\enlargethispage{\baselineskip}


Comparing the data presented in Figs.~\ref{f-asil} and \ref{f-09}
with observations, it is possible to estimate
characteristics of grains fitting the observations:
$r_{\rm s,\,compact} \approx 2.0\,\mkm$ or
$V_{\rm Be1}/V_{\rm solid}\approx 0.2$ and
$r_{\rm s,\,porous} \approx 0.215\,\mkm$
(last value is obtained
from Eq.~(\ref{xpor}) for ${\cal P}=0.9$).
Hence the same {\rm observational} data can be explained with
particles whose radii differ by a factor of $\sim 9$  and
masses  by a factor of $\sim 8800$~(!).

Using the   observations
from Schegerer et al.~(\cite{schetal06});
van Boekel et al.~(\cite{vb05});
Sargent et al.~(\cite{sar06}); and
Sicilia-Aguilar et al.~(\cite{aurore07}),
we fitted observational and theoretical profiles of the silicate
feature. We choose the best model by minimising the $\chi^2$ criterion.
In all cases, the observational continuum is fitted by a straight
line in the interval $8 - 12\, \mu$m for the sake simplicity.
We refer the reader to Juh\'asz et al.~(\cite{ju08}) for a
detailed discussion of the physical models of continuum.
We made the calculations
for layered spheres with $r_{\rm s,\,compact}=0.1\,\mkm$
consisting of Be1  and MgFeSiO$_4$.


\begin{figure}[ht]
\bc
\resizebox{\hsize}{!}{\includegraphics{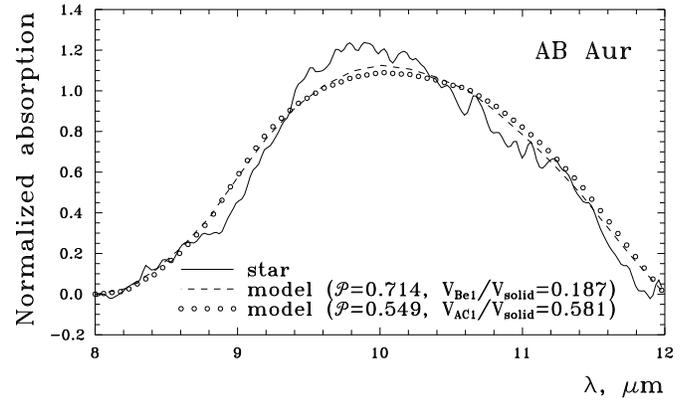}}
\caption{The observational normalised profile of the
10~$\mkm$  feature for the Herbig Ae/Be star AB~Aur
(van Boekel et al.,~\cite{vb05}; continuum is subtracted.)
The best fit profiles are shown for particles containing
amorphous carbon in the form Be1 (dashed line) and AC1
(circles). The values of particle porosity ${\cal P}$ and
fractional amount of carbon $V_{\rm Be1}/V_{\rm solid}$ are indicated.
}\label{ab}\ec
\vspace*{-0.5cm}
\end{figure}

The estimated porosity and amount of carbon in grains producing
the silicate emission feature in protoplanetary discs are collected
in online Table~\ref{t2}.
At this stage, the stars with very pronounced crystalline peaks were
eliminated from consideration.
Table~\ref{t2} includes 47 stars (30 T Tau  and 17 HAeBe stars).
The obtained values of ${\cal P}$, $V_{\rm Be1}/V_{\rm solid}$,
the ratio of masses of carbon to silicate $M_{\rm Be1}/M_{\rm MgFeSiO_4}$
and stellar age (if known) are given. Note that the determination of the
age is {\rm often quite uncertain}
for pre-mainsequence stars (see discussion in
Blondel \& Tjin A Djie,~\cite{blo06}). Therefore, we do not
discuss  possible correlations.

The grain porosity exceeds 0.5 and the average value
of ${\cal P}$ is equal to
$\langle {\cal P}\rangle = 0.64 \pm 0.15$. Such particles
{\rm
resemble aggregates
obtained both experimentally (Kempf et al.,~\cite{kph99}) and theoretically}
(Shen et al.,~\cite{sdj08})\footnote{{\rm See also
discussion in Li et al.~(\cite{llb03}).}}.
The amount of carbon in grains is not very large
(average volume fraction
$\langle V_{\rm Be1}/V_{\rm solid}\rangle = 0.24 \pm 0.10$ and
average mass ratio
$\langle M_{\rm Be1}/M_{\rm MgFeSiO_4}\rangle = 0.19 \pm 0.12$).
{\rm These values increase by a factor of 3 -- 4},
if we replace Be1 by the less absorbing amorphous carbon AC1.
This fact is illustrated in Fig.~\ref{ab}, where the
results {\rm are given for both particle materials.}
Note that the particles containing AC1 are less porous.

\enlargethispage{\baselineskip}

The variation of grain porosity
without significant change of grain mass may explain the behaviour
of silicate emission.
This explanation is an alternative to the commonly used idea of
large grains in protoplanetary discs.
We will be able to decide
between the two hypotheses
after conducting spectropolarimetry
in the 10~$\mkm$  feature because a noticeable
albedo of large grains manifests itself in
polarisation {\rm of the scattered light.}
In this case, we expect unusual behaviour of  polarisation
parameters (especially positional angle) within the feature
profile in comparison with calculated profiles for dichroic
extinction (see, e.g., Henning \& Stognienko,~\cite{hs93};
Prokopjeva \& Il'in,~\cite{ip07}).

\section{Crystalline silicates in composite grains}\label{cry}

\begin{figure}[htb]
\bc
\resizebox{\hsize}{!}{\includegraphics{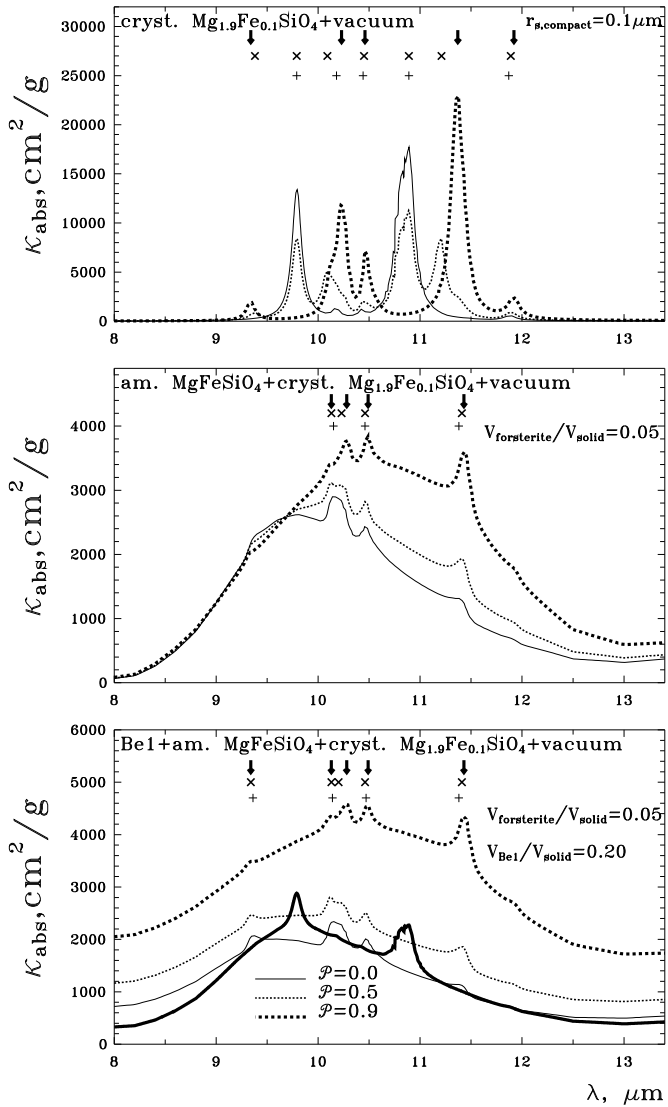}}
\caption{Wavelength dependence of the
mass absorption coefficient for compact and layered
spheres with $r_{\rm s,\,compact}=0.1\,\mkm$.
The particles consist of crystalline olivine and vacuum (upper panel),
amorphous silicate, crystalline olivine, and vacuum (middle panel)  and
amorphous carbon, amorphous silicate, crystalline olivine,
and vacuum (lower panel).
The volume fraction of crystalline olivine and Be1 is indicated.
The position of peaks is marked.
{\rm The thick solid line at the lower panel shows the
mass absorption coefficient for the mixture of compact
grains (external mixing)
with volume fractions of constituents corresponding to
solid curve (internal mixing).}
The effect of variation of crystalline resonances with
particle porosity and composition is illustrated.
}\label{f-cry}\ec
\vspace*{-0.5cm}
\end{figure}

Another interesting problem is the degree of crystallinity
of dust in protoplanetary discs, which is related to the processes
of partial grain evaporation and annealing (e.g., Gail,~\cite{gail04}).
Due to the conversion of amorphous silicates to crystalline minerals,
the particles may consist of different
types of silicates. In order to show the effect of amorphous
silicate matrix and vacuum on resonances
produced by crystalline silicates, we calculated the
feature profiles for composite particles containing Mg-rich
crystalline olivine Mg$_{1.9}$Fe$_{0.1}$SiO$_4$ as a component.
The results are plotted in Fig.~\ref{f-cry} where the upper panel
illustrates the influence of particle porosity on position and strength
of emission peaks. It is seen that variations of spectra are
significant: the shape of the feature changes (cf. Fig.~\ref{f-ob}),
some peaks totally disappear and  new peaks arise.
A very pronounced peak with a maximum near $\lambda= 11.37\,\mu$m is observed
for very porous particles whereas
compact and medium-porous particles  have
resonances  near $\lambda= 10.89\,\mu$m.
The inclusion of crystalline silicates in a composite particle
containing amorphous silicate (middle panel) or
amorphous silicate  and carbon (lower panel) changes the picture.
The resonance near $\lambda \approx 11.4\,\mu$m is clearly seen
at the long-wavelength wing of the feature. Its position slightly
shifts if  the porosity  changes.
Fabian et al.~(\cite{fabetal01})
found that such a peak appeared in the spectra of
strongly-elongated particles.
A double peaked structure around $\lambda= 10.25\,\mu$m arises as well.
This structure was not noticed previously in spectra
of crystalline olivines (H. Mutschke, priv. commun., 2007).
{\rm Note that,  as expected, the mixture of separate constituents
(amorphous silicate, carbon and crystalline silicate,
thick line in lower panel) do not lead to the shift of peaks.}
Further calculations with different materials for wider wavelength range
and detailed comparison with {\it Spitzer} observations
(see Watson et al.,~\cite{wetal07}) might help to resolve the
problem of grain
crystallisation in protoplanetary discs {\rm and to answer the question
whether the crystals occur in ``isolation'' or as part of porous grains.}

\enlargethispage{\baselineskip}


\acknowledgements{
We thank R.~van Boekel,   A.~Schegerer,
B.~Sargent, M.~Honda, and A.~Sicilia-Aguilar
for sending the observational data in tabular form,
H.~Mutschke, V.~Il'in, S.~Korneyev, R.~van Boekel, A.~Juh\'asz
{\rm and the anonymous referee for helpful comments and suggestions.}
The work was partly supported by grants
NSh 8542.2006.2, RNP 2.1.1.2152 and RFBR 07-02-00831
of the Russian Federation.
}


\newpage
\renewcommand{\footnoterule}{}  
\begin{table*}
\caption[]{
Grain porosity and fractional amount of carbon in grains
as derived from fitting the 10~$\mkm$ silicate emission feature.
}    \label{t2}
\bc\begin{tabular}{lcccccccl}
\hline\hline
\noalign{\smallskip}
 Star & Type\footnotemark[1] & ${\cal P}$ &
 $V_{\rm Be1}/V_{\rm solid}$ & $M_{\rm Be1}/M_{\rm MgFeSiO_4}$ &
Age, Myr & Ref.\footnotemark[2] & Comment \\
\noalign{\smallskip}\hline
Tr 37 73-758     &  t  & 0.784      &  0.259    &  0.196     &    1.8  &  1  &                    \\
Tr 37 11-2146    &  t  & 0.705      &  0.217    &  0.156     &    0.9  &  1  &                    \\
Tr 37 11-2037    &  t  & 0.615      &  0.232    &  0.170     &    2.5  &  1  &                    \\
Tr 37 11-2031    &  t  & 0.542      &  0.167    &  0.113     &    2.5  &  1  &                    \\
Tr 37 14-183     &  t  & 0.704      &  0.281    &  0.219     &    0.9  &  1  &                    \\
Tr 37 82-272     &  t  & 0.726      &  0.188    &  0.130     &   10.5  &  1  &                    \\
Tr 37 12-2113    &  t  & 0.650      &  0.186    &  0.128     &    1.1  &  1  &                    \\
Tr 37 13-157     &  t  & 0.804      &  0.250    &  0.187     &    2.4  &  1  &                    \\
Tr 37 91-155     &  t  & 0.976      &  0.320    &  0.264     &    1.7  &  1  &                    \\
Tr 37 54-1547    &  t  & 0.556      &  0.132    &  0.085     &    5.7  &  1  &                    \\
Tr 37 13-1250    &  t  & 0.649      &  0.124    &  0.079     &    3.3  &  1  &                    \\
Tr 37 23-162     &  t  & 0.722      &  0.247    &  0.184     &    6.6  &  1  &                    \\
Tr 37 01-580     &  t  & 0.604      &  0.167    &  0.112     &    8.7  &  1  &                    \\
NGC 7160 DG-481  &  h  & 0.852      &  0.333    &  0.279     &   12.0  &  1  &                    \\
SU Aur           &  t  & 0.731      &  0.131    &  0.085     &    3.9  &  2  &  age from (5)           \\
GW Ori           &  t  & 0.562      &  0.102    &  0.064     &    1.0  &  2  &                    \\
CR Cha           &  t  & 0.471      &  0.107    &  0.067     &    1.0  &  2  &                    \\
Glass I          &  t  & 0.671      &  0.208    &  0.147     &    1.0  &  2  &                    \\
WW Cha           &  t  & 0.539      &  0.194    &  0.135     &    0.3  &  2  &                    \\
SZ 82            &  t  & 0.427      &  0.376    &  0.338     &    1.1  &  2  &                    \\
AS 205S          &  t  & 0.556      &  0.251    &  0.188     &    0.1  &  2  &                    \\
Haro 1-16        &  t  & 0.425      &  0.141    &  0.092     &    0.5  &  2  &                    \\
AK Sco           &  t  & 0.519      &  0.116    &  0.073     &    7.6  &  2  &  age from (5)           \\
FM Tau           &  t  & 0.598      &  0.256    &  0.193     &    2.8  &  3  &  age from (6)           \\
GG Tau A         &  t  & 0.401      &  0.356    &  0.310     &    3.3  &  3  &  age from (7)           \\
GG Tau B         &  t  & 0.841      &  0.342    &  0.292     &    1.6  &  3  &  age from (7)           \\
GM Aur           &  t  & 0.700      &  0.162    &  0.109     &    7.4  &  3  &  age from (7)           \\
IP Tau           &  t  & 0.493      &  0.213    &  0.152     &    4.3  &  3  &  age from (7)           \\
TW Hya           &  t  & 0.756      &  0.158    &  0.105     &   10.0  &  3  &  age from (2)           \\
Hen 3-600 A      &  t  & 0.860      &  0.239    &  0.176     &         &  3  &                    \\
V410 Anon 13     &  t  & 0.651      &  0.398    &  0.370     &         &  3  &                    \\
HD 104237        &  h  & 0.612      &  0.307    &  0.249     &    4.8  &  4  &  age from (5)           \\
HD 142527        &  h  & 0.673      &  0.301    &  0.242     &    1.0  &  4  &                    \\
HD 142666        &  h  & 0.680      &  0.268    &  0.205     &    4.4  &  4  &  age from (5)           \\
HD 144432        &  h  & 0.440      &  0.123    &  0.078     &    6.65 &  4  &  age from (5)           \\
HD 144668        &  h  & 0.832      &  0.504    &  0.569     &    0.5  &  4  &                    \\
HD 150193        &  h  & 0.398      &  0.180    &  0.124     &    2.6  &  4  &  age from (5)           \\
HD 163296        &  h  & 0.537      &  0.209    &  0.148     &    6.0  &  4  &  age from (5)           \\
HD 179218        &  h  & 0.948      &  0.292    &  0.231     &    1.3  &  4  &                    \\
HD 245185        &  h  & 0.704      &  0.149    &  0.098     &    3.3  &  4  &  age from (5)           \\
HD 36112         &  h  & 0.431      &  0.122    &  0.078     &    4.4  &  4  &  age from (5)           \\
HD 37357         &  h  & 0.480      &  0.206    &  0.146     &   10.0  &  4  &                    \\
HD 37806         &  h  & 0.732      &  0.405    &  0.382     &    0.8  &  4  &                    \\
AB Aur           &  h  & 0.714      &  0.187    &  0.129     &    4.8  &  4  &  age from (7)           \\
HK Ori           &  h  & 0.551      &  0.400    &  0.374     &    5.8  &  4  &  age from (5)           \\
UX Ori           &  h  & 0.392      &  0.160    &  0.107     &    2.8  &  4  &  age from (5)           \\
V380 Ori         &  h  & 0.694      &  0.493    &  0.545     &    7.4  &  4  &  age from (5)           \\
\noalign{\smallskip}\hline
\noalign{\smallskip}
\multicolumn{7}{@{}l@{}}{$^1$Meaning of symbols: t = T Tau star, h = Herbig Ae/Be star.}\\
\multicolumn{7}{@{}l@{}}{$^3$References for observational data: (1) Sicilia-Aguilar et al.~(\cite{aurore07});
(2) Schegerer et al.~(\cite{schetal06}); (3) Sargent et al.~(\cite{sar06});}\\
\multicolumn{7}{@{}l@{}}{\hspace{0.8mm}  (4) van Boekel et al.~(\cite{vb05});
(5) Blondel \& Tjin A Djie~(\cite{blo06}); (6) G\"udel et al.~(\cite{gu06});
(7) Bertout et al.~(\cite{ber07}).}\\
\end{tabular}\ec
\end{table*}
\end{document}